\documentstyle{article}
\title{\sc ELLIPTIC AND CIRCULAR WORMHOLES}
\author{ Pedro F. Gonz\'alez-D\'{\i}az.\\
Instituto de Matem\'aticas y F\'{\i}sica Fundamental\\
Consejo Superior de Investigaciones Cientificas\\
Serrano 121, 28006 Madrid (SPAIN)\\
}
\date{June 21, 1993}
\begin{document}
\maketitle
\large
\setlength{\baselineskip}{0.5cm}
\vspace{3cm}

Two new exact analytical solutions of the euclidean Einstein equations for a
minimal massless scalar field and negative cosmological constant have been
obtained. These solutions are given in terms of Jacobian elliptic
or circular functions, rather than hyperbolic functions,
connect large asymptotic regions of maximally-symmetric anti-DeSitter
metrics through a microscopic throat,
and correspond to negative definite components of the Ricci tensor.
Therefore, they describe wormhole-like changes of topology driven by nucleation
of
baby universes. The quantum state of such elliptic and circular
wormholes or handles is
discussed in the most interesting inner and asymptotic regions.

\pagebreak

\section{INTRODUCTION}

The choice of the boundary conditions for the quantum state of the universe
must
be done on just the two natural possibilities which exist for positive definite
metrics; i.e. compact metrics or noncompact metrics which are asymptotic to
metrics of maximal symmetry. Choosing only compact metrics to avoid [1] any
boundary for the quantum state of the universe does not make much difference
for the state amplitude with respect to the alternate choice of noncompact
metrics which are disconnected, consist of a compact part with physical
boundary at a given hypersurface and an asymptotically euclidean or
anti-DeSitter
part without any inner boundary, and dominate on any noncompact connected
metric [2,3]. For practical purposes, instead of compact metrics, one could
likewise take noncompact disconnected metrics for the physical quantum state
of the universe. Therefore, metrics which are asymptotically euclidean or
anti-DeSitter may also contribute the path integral describing the state
of the universe. Clearly, when such metrics are endowed with a given inner
boundary which behaves as a microscopic nonzero bridge between two
asymptotic large regions, the metrics would represent contributions from
wormholes.

Wormholes whose maximally-symmetric asymptotic metrics are euclidean space
have already been extensively considered [4]. They comes about as solutions
of the euclidean Einstein equations for gravity coupled to special kinds of
matter fields [5]. When expressed as a Friedmann-Robertson-Walker metric
with scale factor $a$ in terms of the conformal time
$\eta=\int\frac{d\tau}{a}$,
these wormhole spacetimes can be represented by [6]
\[a = M_{n}\cosh^{\frac{2}{n}}(\frac{n}{2}\eta),\]
with $M_{n}$ the throat radius. Solutions exist for $n=1,2$ and $4$; they
correspond to large [7], Tolman-Hawking [8] and Giddings-Strominger [9]
wormholes, respectively. It has been shown that these three solutions
can be related to each other by means of suitable conformal transformations
[4],
and that each of them converts into one of the most important closed
Robertson-Walker universe models with a perfect fluid (respectively,
dust-filled,
radiation-filled, and ultra-stiff matter universe) by analytical continuation
into the Lorentzian region [10]. However, connections bridging
maximally-symmetric
anti-DeSitter asymptotic metrics have not been hitherto studied. It is the
aim of this paper to examine the spacetime characteristics of two of such
connections. These will appear as particular solutions of the euclidean
equations of motion for Hilbert-Einstein gravity minimally coupled to
scalar fields for the case of a negative cosmological constant.

\section{THE CLASSICAL SOLUTIONS}

The euclidean action corresponding to a minimal massless scalar field $\Phi$,
coupled to Hilbert-Einstein gravity with a negative cosmological constant
$\Lambda$, can be written as
\begin{equation}
\[I=-\frac{1}{16\pi G}\int_{M}d^{4}xg^{\frac{1}{2}}(R+2\Lambda)\]
\end{equation}
where $K$ is the trace of the second fundamental form. In a
 Friedmann-Robertson-Walker
minisuperspace, action (1) reduces to
\begin{equation}
I = -\frac{1}{2}\int d\eta
 N(-\frac{a'^{2}}{N^{2}}-a^{2}-\frac{\Lambda}{3}a^{4}+a^{2}\phi '^{2}),
\end{equation}
where $N$ is the lapse function and $\phi=(\frac{4\pi
G}{3})^{\frac{1}{2}}\Phi$.
$'$ denotes differentiation with respect to conformal time.

{}From the equation of motion for $\phi$, we obtain
\begin{equation}
a^{2}\phi '=B_{0}=Const.
\end{equation}

The holding of (3) implies: (i) that action (2) describes a vacuum whose
energy-momentum tensor will be given in terms of the cosmological constant
$\Lambda$,
and (ii) conservation of the momentum conjugate to $\phi$ which we shall
refer to as the scalar field charge $B_{0}$. The first implication stems
inmediately from the Klein-Gordon equation modified by the Robertson-Walker
evolution, i.e.
\begin{equation}
\ddot{\phi}+3H\dot{\phi}=-\frac{\partial V(\phi)}{\partial\phi},
\end{equation}
where the overhead dot means derivative with respect to the Robertson-Walker
time
$\tau$, $V(\phi)$ is the scalar-field potential, and $H=\frac{\dot{a}}{a}$ is
the Hubble parameter. Hence, from (3) and (4) it follows that $\frac{\partial
V(\phi)}{\partial\phi}=0$.

{}From implication (ii) it turns out that the scalar field charge flowing
through
a three-sphere enclosing the origin is given by $2\pi^{2}B_{0}$. As it was
noted by Myers [11], any wormhole-like solution producing handles in large
parent (here, asymptotic anti-DeSitter) universes would then imply quantized
scalar-field charges, so that $B_{0}=nB_{0}^{(min)}$, with the quantum of
charge defined by $\phi\rightarrow\tilde{\phi}=\phi+\frac{B_{0}^{(min)}}{\pi}$,
$\phi$ identified with $\tilde{\phi}$, and $n$ an integer. We note
furthermore that quantization of the charge should be associated not just to
handles
on only one large region, but also with any even number of wormholes connecting
two otherwise disconnected asymptotic large anti-DeSitter regions. We would
like to emphasize that the lorentzian version of such an spacetime would then
admit an $SL(2,C)$ spinor structure [12].

We shall show in what follows that exact analytical wormhole-like solutions
whose maximally-symmetric asymptotic metrics are anti-DeSitter space can be
obtained from the
equations of motion derived from (2) and (3) if we choose particular suitable
values for the charge $B_{0}$.

The equation of motion for $a$ and the Hamiltonian constraint obtained from (2)
and (3)
are
\begin{equation}
a''=a+\frac{2}{3}\Lambda a^{3}+\frac{B_{0}^{2}}{a^{2}}
\end{equation}
\begin{equation}
a'^{2}=a^{2}+\frac{\Lambda}{3}a^{4}-\frac{B_{0}^{2}}{a^{2}}.
\end{equation}

Solutions to (5) and (6) can be easily obtained using the ansatz
$B_{0}^{2}=\frac{2}{3}\Lambda ^{-2}$
and the change of variables
\[a^{2}=w^{-2}-\Lambda^{-1} , \xi=\int wd\eta .\]
We have two solutions
\begin{equation}
a=\Lambda
^{-\frac{1}{2}}(\frac{3^{\frac{1}{2}}}{\cos(2\Lambda^{-\frac{1}{2}}\xi)}-1)^{\frac{1}{2}}
\end{equation}
\begin{equation}
a=\Lambda^{-\frac{1}{2}}(\frac{3^{\frac{1}{2}}}{\sin(2\Lambda^{-\frac{1}{2}}\xi)}-1)^{\frac{1}{2}}.
\end{equation}

For the first of these solutions, we have for the scalar field
\begin{equation}
\phi=-(\frac{2}{3})^{\frac{1}{2}}\eta
-\frac{3^{\frac{1}{4}}i}{3^{\frac{1}{2}}+1}\Pi(\sin^{-1}[2^{\frac{1}{2}}\cos(\Lambda^{-\frac{1}{2}}\xi)],\frac{1}{3^{\frac{1}{2}}+1},\frac{2^{\frac{1}{2}}}{2}) ,
\end{equation}
where
\begin{equation}
\eta=(\frac{3}{4})^{\frac{1}{4}}F(\sin^{-1}[2^{\frac{1}{2}}\sin(\Lambda^{-\frac{1}{2}}\xi)],\frac{2^{\frac{1}{2}}}{2}) ,
\end{equation}
$F$ and $\Pi$ are the elliptic integral of the first and third kind,
respectively [13], and $0<\xi\leq\frac{\pi}{4}\Lambda^{\frac{1}{2}}$. Solution
(7)
can be expressed in terms of the conformal time $\eta$ by using (10) as
\begin{equation}
a=\Lambda^{-\frac{1}{2}}(3^{\frac{1}{2}}nc^{2}[(\frac{4}{3})^{\frac{1}{4}}\eta]-1)^{\frac{1}{2}},
\end{equation}
whose lorentzian continuation is
\begin{equation}
a=\Lambda^{-\frac{1}{2}}(3^{\frac{1}{2}}cn^{2}[(\frac{4}{3})^{\frac{1}{4}}\eta]-1)^{\frac{1}{2}}.
\end{equation}
In Eqns. (11) and (12) $nc$ and $cn$ denote Jacobian elliptic functions [13].

We also obtain $\tau$ as a function of variable $\xi$,
\[\tilde{\tau}\equiv
(\frac{7+48^{\frac{1}{2}}}{12})^{\frac{1}{4}}\Lambda^{\frac{1}{2}}\tau \]
\begin{equation}
=\Pi(\sin^{-1}[\frac{\Lambda^{-\frac{1}{2}}\xi}{(3^{\frac{1}{2}}-1)(\frac{1}{2}+\frac{\sin^{2}\Lambda^{-\frac{1}{2}}\xi}{3^{\frac{1}{2}}-1})^{\frac{1}{2}}}],\frac{12^{\frac{1}{2}}}{3^{\frac{1}{2}}+1},
\frac{2^{\frac{1}{2}}}{(3^{\frac{1}{2}}+1)^{\frac{1}{2}}}).
\end{equation}

We finally note that the Bochner theorem [14] is respected in the sense that
the scalar curvature and the time-time component of the Ricci tensor are both
negative definite
\[R=\frac{12\Lambda^{-2}}{a^{6}}(3nc^{4}T-3^{\frac{1}{2}}nc^{6}T-3^{\frac{1}{2}}nc^{2}T)<0\]
\[R_{0}^{0}=-\frac{3\Lambda^{-2}}{a^{6}}(3^{\frac{1}{2}}nc^{6}T-3nc^{4}T+3^{\frac{1}{2}}nc^{2}T+1)<0 , \]
where $T\equiv (\frac{4}{3})^{\frac{1}{4}}\eta$.

It follows then that the
topological change implied by this solution is allowed, and therefore it can be
interpreted as a wormhole or handle on asymptotically anti-DeSitter spaces.

Let us take now $B_{0}^{2}=\frac{4}{3}\Lambda^{-2}$. In this case the
analytical
solutions to the equations of motion (3) and (5), and Hamiltonian constraint
(6)
are
\begin{equation}
a=\Lambda^{-\frac{1}{2}}(\frac{3}{\cos^{2}\eta}-2)^{\frac{1}{2}}
\end{equation}
\begin{equation}
a=\Lambda^{-\frac{1}{2}}(\frac{3}{\sin^{2}\eta}-2)^{\frac{1}{2}}.
\end{equation}
For the first of these solutions, we have
\begin{equation}
\phi =
-3^{\frac{1}{2}}(\eta-3^{\frac{1}{2}}\tan^{-1}(3^{\frac{1}{2}}\tan\eta)).
\end{equation}
The lorentzian continuation of (14) becomes
\begin{equation}
a=\Lambda^{-\frac{1}{2}}(\frac{3}{\cosh^{2}\eta}-2)^{\frac{1}{2}}.
\end{equation}
In this case, we have for the Robertson-Walker time
\begin{equation}
\tau=\frac{\Lambda^{-\frac{1}{2}}}{3^{\frac{1}{2}}}\Pi(\sin^{-1}[\frac{3^{\frac{1}{2}}\sin\eta}{(1+2\sin^{2}\eta)^{\frac{1}{2}}}],-(\frac{2}{3})^{\frac{1}{2}},1),
\end{equation}

Again in this case the solutions will correspond to an allowed change of
topology as the curvature scalar and the time-time component of the Ricci
tensor
also are both negative definite. Thus, solution (11) represents an elliptic
wormhole and solution (14) a circular wormhole. Moreover, by shifting the
scale factor squared by some given constants [4], one obtains new
solutions, such as $(a^{2}+a(\eta=0)^{2})^{\frac{1}{2}}$ which gives
$(\frac{2}{\Lambda})^{\frac{1}{2}}(3^{\frac{1}{2}}dc^{2}[(\frac{4}{3})^{\frac{1}{4}}\eta]-1)^{\frac{1}{2}}$
for elliptic wormholes or
$\Lambda^{-\frac{1}{2}}(\frac{3}{\cos^{2}\eta}-1)^{\frac{1}{2}}$
for circular wormholes. Of particular interest is the shift
\begin{equation}
a^{(0)}=(a^{2}+\Lambda^{-1})^{\frac{1}{2}}=3^{\frac{1}{4}}\Lambda^{-\frac{1}{2}}nc[(\frac{4}{3})^{\frac{1}{4}}\eta],
\end{equation}
which yields a scale factor that depends linearly on the Jacobian elliptic
function $nc$ (it is easy to see that the lorentzian continuation of (19)
will lead to the same expression as (19), but with the function $cn$ instead
of $nc$.) The functions $nc$ and $cn$ are doubly periodic, with periods $4K$
and $4Ki$, $K$ being the complete elliptic integral of the first kind
$K\equiv F(\frac{\pi}{2},\frac{2^{\frac{1}{2}}}{2})$ [13]. It follows that,
if the wormhole (19) connects points on just $two$, rather than four, parent
large universes, then the scalar field $\phi$ will take values on a closed
path, and the momentum (scalar-field charge) generating shifts of $\phi$
will become quantized. However, if the charge is quantized, solution (11)
becomes a handle on just $one$ asymptotic anti-DeSitter region. Considering
two of such regions leads to a situation which would correspond to a pair
($n=2$) of scalar-field quanta, each with charge
$(\frac{2}{3})^{\frac{1}{2}}\Lambda^{-1}$,
such as it is depicted in Fig. 1, as compared with wormhole (19) for
which the scalar-field charge is quantized so that $n=1$. For $n=1$, the
two large universes are connected by two wormhole tubes, and for $n=2$ they
are disconnected and each possesses a handle. It is worth emphasising that,
since these two elliptic wormhole solutions are periodic both in real and
imaginary conformal time, and so should possess gravitational temperature,
the essential difference between such solutions is that whereas the conformal
time periods are $2K$ and $2Ki$ for spacetime (11), such periods become $4K$
and $4Ki$ for spacetime (19). It then follows that wormhole (19) has a
temp
associate with wormhole (11), i.e. $\frac{1}{2K}$.
A similar situation can be
likewise discussed for the circular wormhole (17) and the metric which is
derived from it according to the shift $(a^{2}+2\Lambda^{-1})^{\frac{1}{2}}$.
In all the cases, baby universes with metrics correspondingly obtained from
lorentzian continuation from the euclidean wormhole metrics are branched off
from the asymptotic anti-DeSitter regions.

\section{QUANTUM DESCRIPTION}

In order to accomplish a quantum description of the elliptic and circular
wormholes considered in the precedent section, one should not introduce any $a$
$priori$
distinction between different values of the parameter $B_{0}$, dealing with the
general case
and leaving any analysis on the quantum state to be established according to
suitable
boundary conditions. For these we take the Hawking-Page boundary conditions
[15], i.e. we shall only consider those solutions of the Wheeler DeWitt
equation that are regular as the three-geometry degenerates and vanish as the
three-metric becomes infinite. The Wheeler DeWitt equation can be obtained
by introducing quantum operators instead of the classical momenta $p_{a}=a'$
of the Hamiltonian constraint (6). This gives
\begin{equation}
l_{p}^{4}\frac{\partial^{2}\Psi}{\partial
a^{2}}-(a^{2}+\frac{\Lambda}{3}a^{4}-\frac{B_{0}^{2}}{a^{2}})\Psi=0,
\end{equation}
where $l_{p}$ denotes Planck length.

Solving rigurously (20) is very difficult, but we can still obtain approximate
solutions in the two regions of most interest, i.e. as $a\rightarrow\infty$
and as $a\rightarrow 0$. The substitution $y=a^{2}$ transforms (20) into
\begin{equation}
l_{p}^{4}y^{2}\frac{\partial^{2}\Psi}{\partial
y^{2}}-\frac{1}{4}(y^{2}+\frac{\Lambda}{3}y^{3}-B_{0}^{2})\Psi=0.
\end{equation}
Then, for very large $a$, one can neglect $B_{0}^{2}$ in (21). Defining
$z=1+\frac{\Lambda}{3}y$ and denoting $l_{p}^{4}\Lambda^{2}$ by the
dimensionless
constant $\lambda^{2}$, (21) reduces to
\[\frac{\partial^{2}\Psi}{\partial z^{2}}-\frac{9}{4}\lambda^{-2}z\Psi=0\]
or
\begin{equation}
\frac{\partial^{2}\Psi}{\partial x^{2}}-x\Psi=0,
\end{equation}
with $z^{3}=\frac{4\lambda^{2}}{9}x^{3}$. The solution of (22) can be given in
terms of
the known Airy functions [13,16].

In order to satisfy boundary conditions at infinity, one should choose the
pair of linearly independent solutions $Ai(-x)$ and $Bi(-x)$, which correspond
to the Bessel function
\begin{equation}
\Psi(z)=z^{\frac{1}{2}}{\it C}_{\frac{1}{3}}(i\lambda^{-1}z^{\frac{3}{2}})
\end{equation}
whenever we take for ${\it C}_{\frac{1}{3}}$ the Hankel function
$H^{(1)}_{\frac{1}{3}}$.
Thus, we have in the limit $a\rightarrow\infty$,
\begin{equation}
\Psi(x)=3^{\frac{1}{6}}(2\lambda)^{\frac{1}{3}}[Ai(-x)-iBi(-x)]\rightarrow\Psi(a)=\frac{1}{2}a^{-\frac{1}{2}}e^{-\frac{2}{3}a^{3}},
\end{equation}
which is valid both for positive and negative $a$. The wave function $\Psi(a)$
for asymptotically anti-DeSitter wormholes will be then exponentially damped
at a rate which is quicker than that for asymptotically euclidean wormholes.

It is worth noticing that the wave equation for a particle of unit mass and
charge $e$ moving along direction $\rho$ in a homogeneous electric field
of intensity $E$ is given by $\varphi ''+\sigma\varphi=0$, where [16] in
natural units ($\hbar=c=G=1$)
\[\sigma=(1+e\rho)(\frac{2E^{3}}{F^{2}})^{\frac{1}{3}},\]
$F$ being the force acted upon the particle, $F=eE$. Thus, comparing the
expression for $\sigma$ with
\[x=(1+\frac{\Lambda}{3}a^{2})(\frac{9}{4\Lambda^{2}})^{\frac{1}{3}},\]
(which is also given in natural units)
we see that the asymptotic dynamics of the considered wormhole is equivalent
to that of a particle of unit mass and charge $e=\frac{\Lambda}{3}a$, moving
in a homogeneous field of intensity $E=\frac{1}{8a^{2}}$.

For very small $a$, we obtain from (21)
\begin{equation}
y^{2}\frac{\partial^{2}\Psi}{\partial
y^{2}}-(\frac{y^{2}}{4l_{p}^{4}}-\frac{\bar{B}_{0}^{2}}{4})\Psi=0,
\end{equation}
in which $\bar{B}_{0}^{2}=\frac{B_{0}^{2}}{l_{p}^{4}}$. Solutions to (25) are
given in terms of the modified Bessel functions [13]
\begin{equation}
\Psi(a)=a{\it
T}_{\frac{1}{2}(1-\bar{B}_{0}^{2})^{\frac{1}{2}}}(\frac{a^{2}}{2l_{p}^{2}}).
\end{equation}
The behaviour of $\Psi(a)$ will be regular at $a\rightarrow 0$ if we choose
${\it T}\equiv{\it I}$ [13], since then $\Psi(a)$ tends to zero as one
approaches
$a=0$. Note that even though the solution oscillates for ${\it B}_{0}>1$, the
presence of a factor $a$ will make such oscillations to damp as the
three-geometry
degenerates. This situation is maintained even when we would choose an operator
ordering other than $p=0$. For $p=1$ (which is the case discussed by Hawking
and Page in [15]) we obtain a solution which still vanishes as $a\rightarrow
0$.
In any event, one always can transform to new coordinates in minisuperspace in
terms of which the Wheeler DeWitt equation for small $a$ becomes the wave
equations for two harmonic oscillators with oposite sign of energy, which has
a fully regular behaviour for degenerating three-geometries. The conclusion
is thereby obtained that although asymptotically anti-DeSitter and euclidean
wormholes have quite a different asymptotic behaviour, they behave in a
indistinguishable fashion at the smallest quantum values of the scale factor.

\vspace{1.3cm}

$Acknowledgements$. The author is grateful to CAICYT for support under
Research Project N' PB91-0052 and to CSIC for an Accion Especial. I would
also like to thank L.J. Garay for conversations.

\pagebreak

\noindent\section*{References}
\begin{description}
\item [1] S.W. Hawking, in {\it Astrophysical Cosmology}, Pontificae
Scientiarum
Scripta Varia, Vatican City, eds. H.A. Bruck, G.V. Coyne and M.S. Longair,
1982.
\item [2] J.B. Hartle and S.W. Hawking, Phys. Rev. D28, 2960 (1983).
\item [3] S.W. Hawking, Nucl. Phys. B239, 257 (1983).
\item [4] S.W. Hawking, Phys. Rev. D37, 904 (1988); P.F. Gonz\'alez-D\'{\i}az,
Phys. Rev. D40, 4184 (1989); Int. J. Mod. Phys. A7, 2355 (1992).
\item [5] A. Zhuk, Phys. Rev. D45, 1192 (1992).
\item [6] P.F. Gonz\'alez-D\'{\i}az, Phys. Lett. B261, 357 (1991).
\item [7] P.F. Gonz\'alez-D\'{\i}az, Nuovo Cim. B106, 335 (1991); Mod. Phys.
Lett. A5, 2305 (1990).
\item [8] J.J. Halliwell and R. Laflamme, Class. Quant. Grav. 6, 1839 (1989).
\item [9] S.B. Giddings and A. Strominger, Nucl. Phys. B306, 890 (1988).
\item [10] A. Zhuk, Phys. Lett. A176, 176 (1993).
\item [11] R.C. Myers, Nucl. Phys. B323, 225 (1989).
\item [12] G.W. Gibbons and S.W. Hawking, Commun. Math. Phys. 148, 1 (1992).
\item [13] M. Abramowitz and A. Stegun, {\it Handbook of Mathematical
Functions} (Dover, New York, 1972).
\item [14] M. Bochner, Bull. Am. Math. Soc. 52, 776 (1946).
\item [15] S.W. Hawking and D.N. Page, Phys. Rev. D42, 2655 (1990).
\item [16] L.D. Landau and E.M. Lifshitz, {\it Mec\'anica Cu\'antica
No-Relativista}. Vol. 3. (Ed. Revert\'e, Barcelona, 1967).

\end{description}

\pagebreak

{\bf Legend for Figure}

\vspace{1cm}

Fig. 1.- Pictorial  representation of the two elliptic wormhole instantons
as interpreted in terms of the two first excited states of the scalar-field
charge. The ground state $n=0$ would be chargeless and correspond to just
topologically-featureless asymptotic anti-DeSitter space, without any
wormhole or handle.

\end{document}